Reply to
Comment on "Doppler signature in electrodynamic retarded potentials,"
Giovanni Perosa, Simone Di Mitri, William A. Barletta, Fulvio Parmigiani

Calin Galeriu (CG) has commented [1] about our manuscript, "Doppler signature in electrodynamic retarded potentials," [2] that must be properly addressed, amending at the same time of notation errors [3] to clarify some formal aspects which were discussed and criticized in C. Galeriu's comments.
The same author (CG) has published three papers concerning the origin of the Doppler term, v/c, in the Liénard-Wiechert (L-W) potentials [4-6]. Unfortunately, these papers were not cited in our paper, as we became aware of them after the comment in ref. 1. We apologize for this oversight.

In particular, the statement on page 2 of ref. 2, " Instead of using the well-known identity, the authors derive it, in a convoluted manner, with the help of a Fourier representation of the Dirac delta function." Referring to the use of a property of the Dirac distribution, CG is incorrect and reflects neither in the contextual nor formal aspects of what the authors wrote in ref. 2. Reference 2 clearly states that an alternative route would be the use of this "well-known" identity. However, we preferred to avoid the use of such machinery. One might prefer an alternative choice, but it is certainly not a source of error, either formal or conceptual.

Concerning his comment on page 3 in ref. 1, we agree to disagree. Despite the very clear and proficient explanation on the use of the Dirac distribution that the author gives, we still believe that our approach brings out the physics behind the origin of the L-W potentials.

Finally, at page 4, the expression by Kapoulitsas is not simply mentioned in the Appendix, but is derived. We thank the author of ref. 1 for bringing this reference to our attention. That said, we realized that some of the issues raised by the author require a clarification. Let us indicate with $\phi(\mathbf{s}, t)$, the phase of a spherical wave,

$$\phi(\mathbf{s},t) = \mathbf{k} \cdot \mathbf{s} - \omega t, \qquad (1)$$

This notation may lead to possible confusion, and for this we apologize. Therefore, we have sent the journal an errata corrige, attached here as a note, where we simply change the notation of the variable. Noteworthy, with this clarification, the rest calculations and the conclusions remain the same as in ref. 1. It is a given that the condition of retarded time can be found as,

$$\phi(\mathbf{r},t) = \phi(\mathbf{r}_O(t_r), t_r) \qquad (2)$$

with the quantities as named after Figure 1 of ref. 1.

This condition translates into the requirement of having equal phase at the source and at the position P.

Consistently with the notation in (1), the second relevant consequence of the motion of the source is the *Doppler shift* of the wave frequency,

$$\omega' = -\frac{d\phi}{dt} = -\mathbf{k} \cdot \frac{d\mathbf{s}(t)}{dt} + \omega = \omega\left(1 - \frac{\hat{\mathbf{k}} \cdot \mathbf{v}_s}{c}\right) = \frac{\omega}{D(\mathbf{v}_s, \hat{\mathbf{k}})} \tag{3}$$

$$\mathbf{v}_s = \frac{d\mathbf{s}(t)}{dt}, \quad D(\mathbf{v}_s, \hat{\mathbf{k}}) = \left(\frac{c}{c - \hat{\mathbf{k}} \cdot \mathbf{v}_s}\right) \tag{4}$$

Calculation hereafter remains the same. Eq. (14) in the paper is consistent with definition (1).

We emphasize here that the analyses by CG are purely centered on formal physics-mathematics issues. In particular, although not explicitly stated, they highlight the fact that the axioms underlying Euclidean geometry, such as the definition of a point, have no counterpart in physics. This lack is well reflected in CG's works which correctly point out that the "*Dirac delta function used in electrodynamics must be the one that obeys the weak definition, non-zero in an infinitesimal neighborhood, and not the one obeying the strong definition, non-zero in a point*" [6].

We agree that these physics-mathematics formalisms are quite important in deriving the L-W potentials. However, we emphasize that the derivation of the L-W potentials, as reported in many textbooks and papers, is a case among those arising from physical theories, including classical electrodynamics, that is built on the base of the Euclidean geometry. Of course, this question is fundamental and in CG's work its formal consequences and the related corrections represent an important achievement. This includes the mathematical solution suggested by CG in which charged particles can be represented by "infinitesimal segments" along their own worldlines in the Minkowski space. thereby avoiding the axiomatic definition of point particle given by Euclidean geometry, but assuming a weak Dirac-$\delta$ function.

In our work [2], we derive the L-W potentials starting from the conjecture that they are a consequence of a Doppler effect, as suggested by the characteristic correction factor (*v/c*) present in their analytical formulation. In this framework, our derivation of the L-W potentials is based on physical phenomenology. An important consequence results from the fact that a Doppler effect does not depend on the size of the source and thus on the geometry adopted. Instead, the origin of the (*v/c*) term in the L-W potential is a factor that takes into account the change in phase velocity of the signal coming from a moving source. In conclusion, we believe that the derivation of the L-W potentials as reported in ref. 2 is formally correct at it makes evident the physical phenomenology from which the L-W potentials in electrodynamics originate.

Finally, we stress that the purpose of our derivation is practical. Indeed, L-W potentials are derived in a framework that is not Lorentz-covariant. Our idea is to provide expressions in a preferred reference system, i.e. the laboratory frame, as it is usually done in accelerator physics, that can provide a recipe for operational implementations of the formulas. For this additional reason, we consider the way in which the derivation is presented to be relevant.

References

1- Calin Galeriu, arXiv:2302.13404 [physics.class-ph] (2023).

2- Giovanni Perosa, Simone Di Mitri, William A. Barletta, Fulvio Parmigiani, Physics Open **14**, 100136 (2023).

3- Corrigendum ref. 2
   i) "David J. Griffith" should be "David J. Griffiths" along the text (page 1, second column and references).
   ii) In Eq. (5), "$e^{i\omega t}$" should be "$e^{-i\omega t}$".
   iii) "proportional to δ(t+r/c)" instead of "proportional to (t+r/c)" at the end of the first column at pag. 2 ref. 2.
   iv) In Eq. (18), "$\hat{\mathbf{k}}$" should be "$\hat{\mathbf{R}}$".

4- Călin Galeriu, The geometrical origin of the Doppler factor in the Liénard-Wiechert potentials, *Eur. J. Phys.* **42** 055204 (2021). **DOI** 10.1088/1361-6404/ac1147

5- Călin Galeriu, A derivation of the Doppler factor in the Liénard-Wiechert, *Eur. J. Phys.* **42** 055203 (2021). **DOI** 10.1088/1361-6404/ac1145

6- Călin Galeriu, The algebraic origin of the Doppler factor in the Liénard–Wiechert potentials, *Eur. J. Phys.* **44** 035203 (2023) **DOI** 10.1088/1361-6404/acc991